\documentclass [11pt, letterpaper]{article}
\pdfoutput=1 
\usepackage{fullpage}
\usepackage{amsmath}
\usepackage{amssymb}
\usepackage{graphicx}
\usepackage{braket}
\usepackage{mathrsfs}
\usepackage{wrapfig}
\usepackage{boxedminipage}
\usepackage{epsfig}
\usepackage{setspace}
\usepackage{subfigure}
\usepackage{float}
\usepackage{hyperref}
\usepackage{enumerate}
\usepackage{cite}
\usepackage{bbold}
\usepackage[font=footnotesize,labelfont=rm]{caption}

\begin{document}

\begin{flushright}
{\tt arXiv:1905.$\_\,\_\,\_\,\_\,\_$}
\end{flushright}

{\flushleft\vskip-1.35cm\vbox{\includegraphics[width=1.25in]{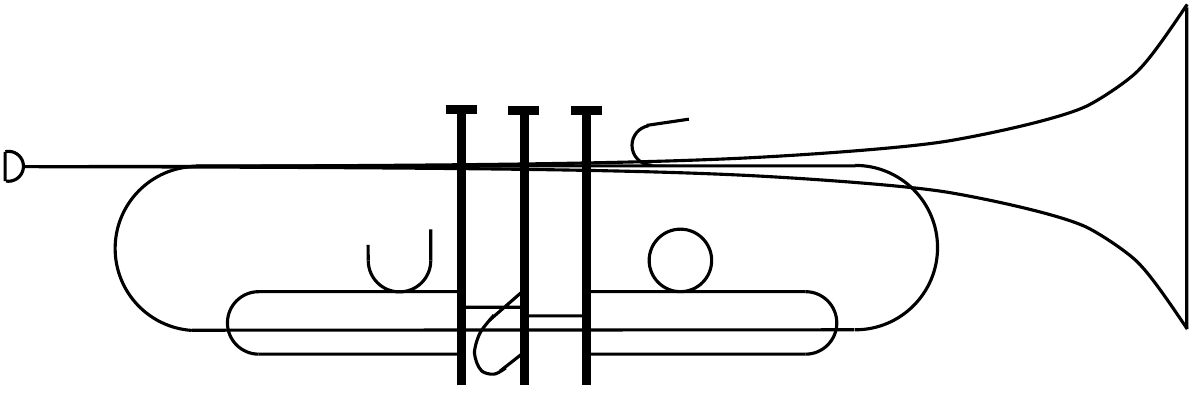}}}

\bigskip
\bigskip

\bigskip
\bigskip
\bigskip
\bigskip

\begin{center} 

{\Large\bf   Specific Heats and Schottky Peaks}

\bigskip

{\Large\bf for}

\bigskip

{\Large\bf  Black Holes in Extended Thermodynamics}

\end{center}

\bigskip \bigskip \bigskip \bigskip

\centerline{\bf Clifford V. Johnson}

\bigskip
\bigskip
\bigskip

\centerline{\it Department of Physics and Astronomy }
\centerline{\it University of
Southern California}
\centerline{\it Los Angeles, CA 90089-0484, U.S.A.}

\bigskip

\centerline{\small {\tt johnson1}  [at] usc [dot] edu}

\bigskip
\bigskip


\begin{abstract} 
\noindent 
In the extended thermodynamics of black holes, there is  a dynamical pressure and  its conjugate volume. The phase structure of many of these  black holes has been studied a great deal and shown to give close analogues of the phase structure of various ordinary  matter systems. However, we point out that  the most studied black holes in this framework, such as Schwarzschild--AdS and Reissner--Nordstr\"{o}m--AdS,  and various analogues in higher--derivative gravity, do not have the type of elementary degrees of freedom that play a central role in the classic models of matter. This is because they have vanishing specific heat at constant volume,~$C_V$.  As examples with non--vanishing $C_V$, the Kerr--AdS and STU--AdS black holes do have such degrees of freedom, and a study of $C_V(T)$ reveals  Schottky--like behaviour suggestive of a finite window of energy excitations. This intriguing physics may have useful applications in fields such as holographic duality, quantum information, and beyond.

\end{abstract}

\pagenumbering{gobble}

\newpage 

\pagenumbering{arabic}

\baselineskip=18pt 
\setcounter{footnote}{0}

\section{Introduction}
\label{sec:introduction}

The framework of black hole thermodynamics\cite{Bekenstein:1973ur,Bekenstein:1974ax,Hawking:1974sw,Hawking:1976de} can be  extended\cite{Kastor:2009wy,Wang:2006eb,Sekiwa:2006qj,LarranagaRubio:2007ut} to allow for not just descriptions of the core thermodynamic quantities of thermal energy $U$, temperature $T$, and entropy~$S$, but also a dynamical pressure\footnote{See also earlier work in refs.~\cite{Henneaux:1984ji,Henneaux:1989zc,Teitelboim:1985dp}.} $p$ and its conjugate, the volume $V$.  The pressure enters the framework by  allowing the cosmological  constant $\Lambda$ of the theory to be dynamical: $p=-\Lambda/8\pi G$. The case of negative $\Lambda$, where the pressure is positive, most readily leads to discussions of stable thermodynamics, and indeed there is a large literature of studying the extended thermodynamics of various black hole solutions. It has been noticed (starting with the casting of the observations of refs.~\cite{Chamblin:1999tk,Chamblin:1999hg} into this framework\cite{Kubiznak:2012wp}) that the thermodynamic phase structure often resembles that of models of familiar interacting matter systems such as the van der Waals gas, {\it etc.} For a review, see ref.\cite{Kubiznak:2016qmn}. 
It is rather mysterious that the phase structure (including  critical exponents) maps so readily to that of non--gravitational systems, and unravelling that mystery may well lead to useful insights into the nature of  gravity, especially when combined with quantum mechanics. It may also have useful applications as a source of non--trivial thermodynamic models with readily accessible equations of state for non--gravitational applications, as suggested in refs.~\cite{Johnson:2017hxu,Johnson:2017ood}\footnote{These observations have, {\it \'a priori}, nothing to do with holographic duality such as AdS/CFT\cite{Maldacena:1997re,Witten:1998qj,Gubser:1998bc}. In that context, being able to change $p$ turns out to have an interpretation in terms of moving on the space of theories, changing the degrees of freedom, the precise dictionary depending upon the context (see {\it e.g.,} refs.~\cite{Johnson:2014yja,Kastor:2014dra,Karch:2015rpa,Johnson:2018amj}). The holographic interpretation of the thermodynamic volume $V$ in the dual field theories is a little more mysterious (it is not the volume of a vessel that contains the field theory any more than the pressure is the thermodynamic pressure in the field theory), but see a companion work to this paper\cite{metoappear} for some progress.}.

More insight might be gained by considering the nature of the degrees of freedom that are in play in these systems, and that is what we do in this note, following some basic lessons from the history of thermodynamics and quantum physics. The point is that a study of the specific heats of a substance can give some insight into the available degrees of freedom (classical or quantum mechanical). Moreover, the contrasting nature of the specific heats at constant pressure $C_p$ and constant volume $C_V$ can provide useful clues. This goes back to the early models of gases provided by Einstein\cite{Einstein1907}, and  Debye\cite{Debye1912}, that allow the derivation of  key measurable thermodynamic observables like $C_V$, starting  from simple statistical mechanical considerations of the basic degrees of freedom (for a review see {\it e.g.,} refs.~\cite{mandl1995statistical,rosser1982introduction,kittel1996introduction}).  Our key point here is that the behaviour of the specific heat {\it as a function of temperature} is an important, if coarse--grained, diagnostic of the degrees of freedom of a system. Interestingly, this behaviour seems to be infrequently examined in the extended thermodynamics literature for black holes, and our point  is that there is information to be learned from it. 

The core elements we need are all readily available in the black hole physics. The usual specific heat of black holes in ordinary thermodynamics, $C=T\partial S/\partial T$, becomes $C_p$ in the extended framework. For  Schwarzschild black holes that are small compared to the scale $\ell$ set by the cosmological constant ($\Lambda=-3/\ell^2$ in $D=4$),~$C_p$ is negative. This is the familiar\cite{Hawking:1974sw} result that such black holes wish to undergo runaway evaporation by radiating energy at a temperature that increases the smaller they get. But for large black holes $C_p$ is positive, and such stable black holes are central to 
a great deal of applications in theoretical physics through holographic studies\cite{Witten:1998zw} (using traditional {\it i.e.,} unextended black hole thermodynamics---see refs.\cite{Aharony:1999t} for  reviews), the aforementioned studies in extended thermodynamics, and even  combinations of the two involving quantum information\cite{Johnson:2018amj,Johnson:2018bma} or  the construction of holographic heat engines and refrigerators that use black holes as working substances\cite{Johnson:2014yja}.

Let us recall some simple thermodynamics. The first law is $dU=TdS-pdV$. Coupling the system to an external  heat bath, one can try to raise the temperature (and hence the internal energy) of the system. The specific heat $C$ is a measure of how much heat is needed to do this. The available degrees of freedom  reveal themselves if  they are present to store this energy in their  excitation. At different temperatures, different degrees of freedom can become available, and so~$C(T)$ is a useful function for keeping track of them. For example, in a diatomic gas, sudden increases in  $C(T)$ around  certain temperatures reveals the energies where the constituent molecules' rotational and (later) vibrational degrees of freedom become accessible.  Crucially,  the specific heat at constant pressure, $C_p$ also takes into account the fact that the system must  do work in order to increase the volume of the system, and so generically it is larger than the specific heat at constant volume, $C_V$. In a sense, $C_V$ is a more direct measure of the available degrees of freedom, at least 
the traditional 
ones  that can be excited without making volume changes. Those are certainly the traditional ones we deal with in ordinary matter made of atoms and molecules. 

It is notable therefore that  the Schwarzschild and Reissner--Nordstr\"{o}m black holes (and  various other simple static black hole solutions of the pure Einstein gravity or higher--derivative generalizations thereof), have $C_V=0$ in extended thermodynamics\footnote{This was first pointed out in this language in ref.~\cite{Dolan:2010ha}.}.  We conclude that {\it this implies that they do not posses such degrees of freedom}. So while the phase structure of these black holes are strikingly similar to that of ordinary matter systems, in this regard they are pointedly different.

There are black holes that have $C_V\neq0$. The obvious pure gravity case to consider is a rotating black hole, the Kerr--AdS solution. The extended thermodynamics and phase structure was worked out in refs.\cite{Cvetic:2010jb,Dolan:2011xt}, extending the computations of ref.\cite{Caldarelli:1999xj}. Since the entropy  and the thermodynamic volume  turn out to be independent functions in that case, this implies $C_V\neq0$.  While this has been noticed before, the temperature dependences of the specific heats $C_p$ and~$C_V$ remain  unexamined in the literature so far,  but can be straightforwardly extracted.  They are difficult to write explicitly, because of the complexity of the equations of state. We therefore explore them numerically, and exhibit some of the key behaviour in section~\ref{sec:kerr-black-holes}.

Since rotation sometimes adds, for some tastes,  an unnecessary interpretational complication (the spacetime at infinity is not static), we also examine the family of ``STU'' black holes (in AdS) as another primary exhibit. These are charged black holes with the same asymptotic geometry as Schwarzschild or Reissner--Nordstr\"{o}m, but coupled to a particular family of scalars\footnote{They have a natural interpretation in terms of spinning D--branes in string theory\cite{Chamblin:1999tk,Cvetic:1999rb}. In fact the Reissner--Nordstr\"{o}m black hole is a special symmetric arrangement of charges/spins that decouples the scalar sector entirely.}.  For these we again explore (in section~\ref{sec:stu-black-holes}) the specific heats in some sample regions of parameter space\footnote{While the complete phase diagram of these black holes has not been fully worked out, we explain why the unknown details cannot alter our key observation about the overall nature of $C_V(T)$.}, finding striking  features similar to that which we observed for Kerr--AdS. 

These are the striking features: For both Kerr and STU, the  function $C_V(T)$, on general grounds, is peaked at some intermediate $T$, {\it falling  to zero at large~$T$}. The temperature scale at which the peak occurs is set by the angular momentum, $J$, or the charge, $Q$, as is its height. This is reminiscent of a Schottky peak (a.k.a., a Schottky anomaly) in studies of ordinary matter\cite{rosser1982introduction,mandl1995statistical,kittel1996introduction}: If there is a sector of the system with  some degrees of freedom  that afford a finite energy window (they could be, {\it e.g.}, impurities in a condensed matter system, hyperfine splitting resulting in some finite set of energy levels, {\it etc.}), then they will make their presence known in $C_V(T)$ as peaks. This is because  at high enough temperatures the energies can be populated, giving a rise in $C_V(T)$, but as $T$ increases beyond a certain point there are no new energy levels to populate, and so the contribution to the specific heat from that sector decreases and eventually falls to zero at large~$T$. 

Our observation  here is that since the entire $C_V(T)$ vanishes at high $T$, ({\it i.e.}, there are no contributions from an unrestricted  continuum or discretuum of available energies as in ordinary matter) these black holes have, {\it in this extended thermodynamic context},  a highly  restricted set of (traditional) degrees of freedom! 
There are some immediate and intriguing consequences of this. The first is that the interesting phase structure that black holes have in extended thermodynamics, although it is similar to that of ordinary matter systems, {\it must have its origins in very different kinds of microscopic physics}. The ordinary translational, rotational, and vibrational degrees of freedom would contribute to $C_V(T)$ at all non--zero $T$, but that is not what is seen here. The second is that there ought to be interesting applications of the black hole dynamics that exploits only our restricted degrees of freedom {\it i.e.,} involving processes with non--trivial flows of heat that are isochoric. These could be useful in a wide range of contexts where black holes are relevant, from holography to quantum information. We will discuss implications further in section~\ref{sec:discussion}.

\section{Reissner--Nordstr\"{o}m--AdS Black Holes}
\label{sec:rn-black-holes}

Let us study the  Reissner--Nordstr\"{o}m--AdS black hole. While the physics we will present is mostly well known\cite{Chamblin:1999tk}, it is useful to review it here for comparison with what is to follow. We will work in four dimensions for  clarity of presentation, but other dimensions will have the same key features. The metric  is:
\begin{equation}
\label{eq:RN-AdS-solution}
ds^2 = -F(R)dt^2+F(R)^{-1}dR^2+R^2(d\theta^2+\sin^2\theta d\phi^2)\ ,\quad{\rm with}\quad
\Delta(R)	= 1-\frac{2m}{R}+\frac{Q^2}{R^2}+\frac{R^2}{\ell^2}\ ,
\end{equation}
where  $Q$ is the physical  charge, the length  $\ell$ is related to the cosmological constant by $\Lambda=-3/\ell^2$, and $0\le R \le \infty$, and $(\theta,\phi)$ are standard coordinates on a round $S^2$. The horizon is at $R_+$, the largest solution of $F(R_+)=0$, associating a mass with a given horizon  radius:
\begin{equation}
m(R_+)=\frac{R_+}{2}\left(1+\frac{Q^2}{R_+^2}+\frac{R_+^2}{\ell^2}\right)\ .
\end{equation}
The traditional thermodynamics identifies  the  internal energy $U$ with the mass  and gives the temperature $T$ and the entropy $S$ as:
\begin{eqnarray}
T=\frac{F^\prime(R_+)}{4\pi} = \frac{1}{4\pi }\left(\frac{1}{R_+}+\frac{3 R_+}{\ell^2}-\frac{Q^2}{R_+^3}\right) \ ,\qquad S=\frac{A}{4}=\pi R_+^2\ .
\end{eqnarray}
However, the extended thermodynamics associates the mass with the enthalpy\cite{Kastor:2009wy} $H\equiv U+pV=m(R_+)$, and the pressure is  $p=-\Lambda/8\pi=3/8\pi\ell^2$. Now  since the first law is $dH(S,p)=TdS+Vdp$, 
\begin{wrapfigure}{r}{0.48\textwidth}
\centering
\includegraphics[width=0.48\textwidth]{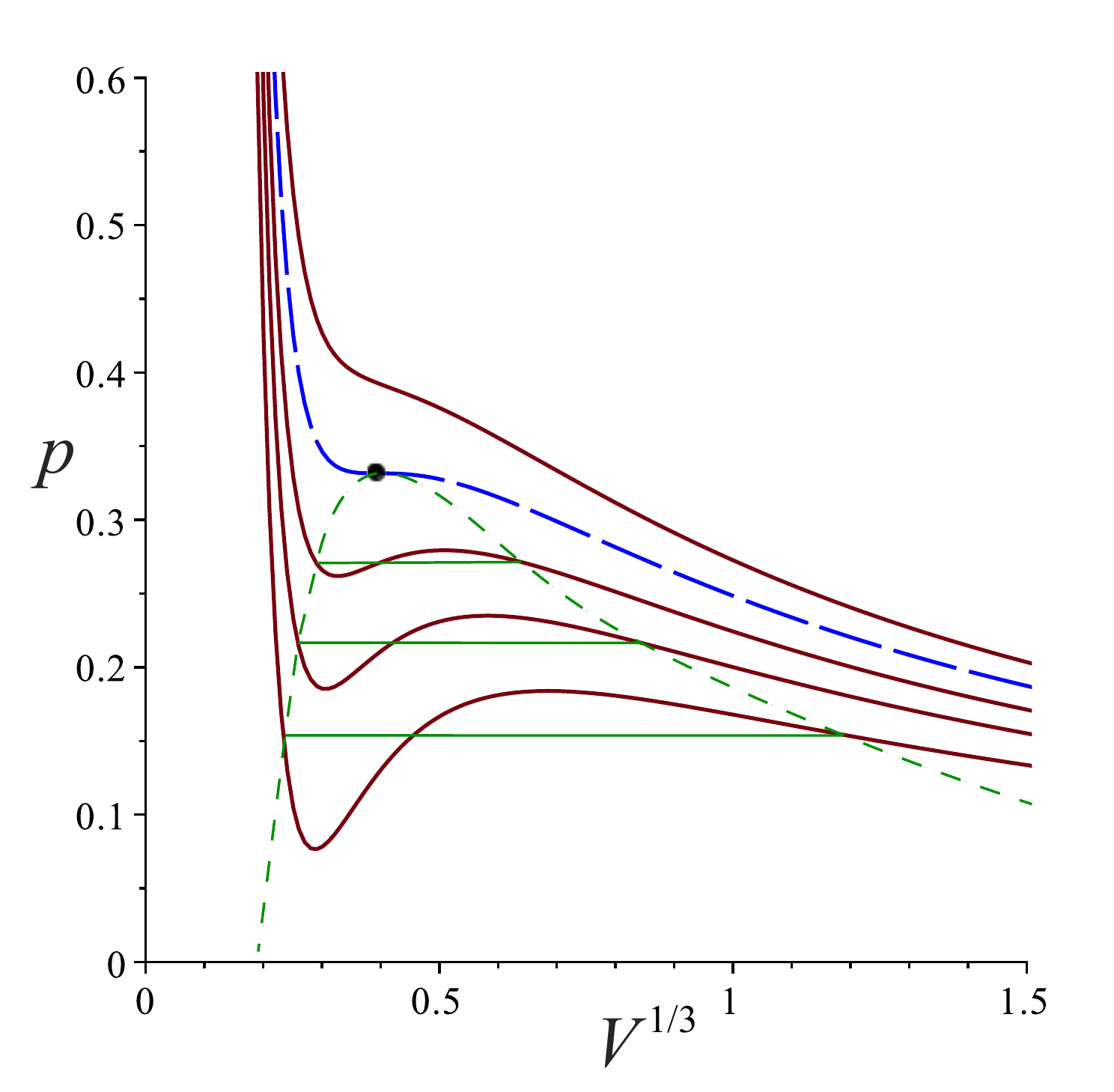}
\caption{\label{fig:pv-diagram} Sample isotherms for the Reissner--Nordstr\"{o}m case. The critical isotherm at $T_c$ is long--dashed (blue). See text for explanation.}
\end{wrapfigure}
we have $T=\partial H/\partial S|_p$ and $V=\partial H/\partial p|_S$ and so, defining $\alpha\equiv{3/4\pi}$, we have:
\begin{equation}
\label{eq:temperature-and-volume-RN}
T=\frac{1}{4\pi }\left(8\pi p (\alpha V)^{1/3}  +\frac{1}{(\alpha V)^{1/3}}-\frac{Q^2}{\alpha V}\right)\ , 
\end{equation}
and 
\begin{equation}
 V=\frac{4\pi}{3} R_+^3\ .
\end{equation}
A most pleasing feature of this system is the fact\cite{Chamblin:1999tk} that  the equation of state $T(p,V)$ includes large and small black holes that  are stable, with positive $C_p$. There are intermediate sized holes that have negative $C_p$ (unstable) but above a critical temperature $T_c=1/(3\sqrt{6}\pi Q)$ (at which $p_c=1/(96\pi Q^2)$ and $V_c=8\sqrt{6}\pi Q^3$) they are not present. See figure~\ref{fig:pv-diagram}, where sign($C_p$) is $-$sign(slope) of an isotherm. 

Below this second order critical point the intermediate ($C_p<0$) black holes are present in the equation of state, but a thermodynamic analysis shows that there is a first order phase transition that skips over this sector, jumping  between the small and large branch (approximately indicated by the hand--drawn (green) horizontal lines in figure~\ref{fig:pv-diagram}). This is the same van der Waals structure that models the liquid--gas phase phase diagram.  In the figure, the (green) short--dashed lines indicate the region of volumes that are excluded by the first order transition at a given temperature. The second order point is shown as a circle on the critical isotherm at $T_c$. Above it, there are no first order subtleties, a point we will use in what is to follow.

Returning to the discussion of specific heats, the key observation here for our purposes is that the entropy  $S$ and volume $V$ are both geometrical, and hence depend on the same variable, the horizon radius $R_+$. The result is that changes in entropy cannot be made without changing the volume, or {\it vice--versa}.  This  is equivalent to saying\cite{Dolan:2010ha} that this class of black holes has the {\it constant volume} specific heat $C_V=0$. On the other hand,  $C_p$ can be computed in terms of the other thermodynamic variables in closed form. For example, writing it as a function on the $p{-}V$ plane\cite{Dolan:2010ha,Kubiznak:2012wp}:
   \begin{equation}
   \label{eq:Cp-4Q}
   C_p(p,V)=2\pi \left( \alpha V \right)^{2/3} 
  \left(  \frac{ 8\pi p
 \left( \alpha V \right) ^{4/3}-{Q}^{2}+ \left( \alpha V \right) ^{2
/3}  }{8 \pi p \left( \alpha V \right) ^{4/3}+3 {Q}^{2}-
 \left( \alpha V \right) ^{2/3}} \right)\ , \quad {\rm where}\quad \alpha=\frac{3}{4\pi} \ .
   \end{equation}
\begin{wrapfigure}{r}{0.5\textwidth}
\centering
\includegraphics[width=0.5\textwidth]{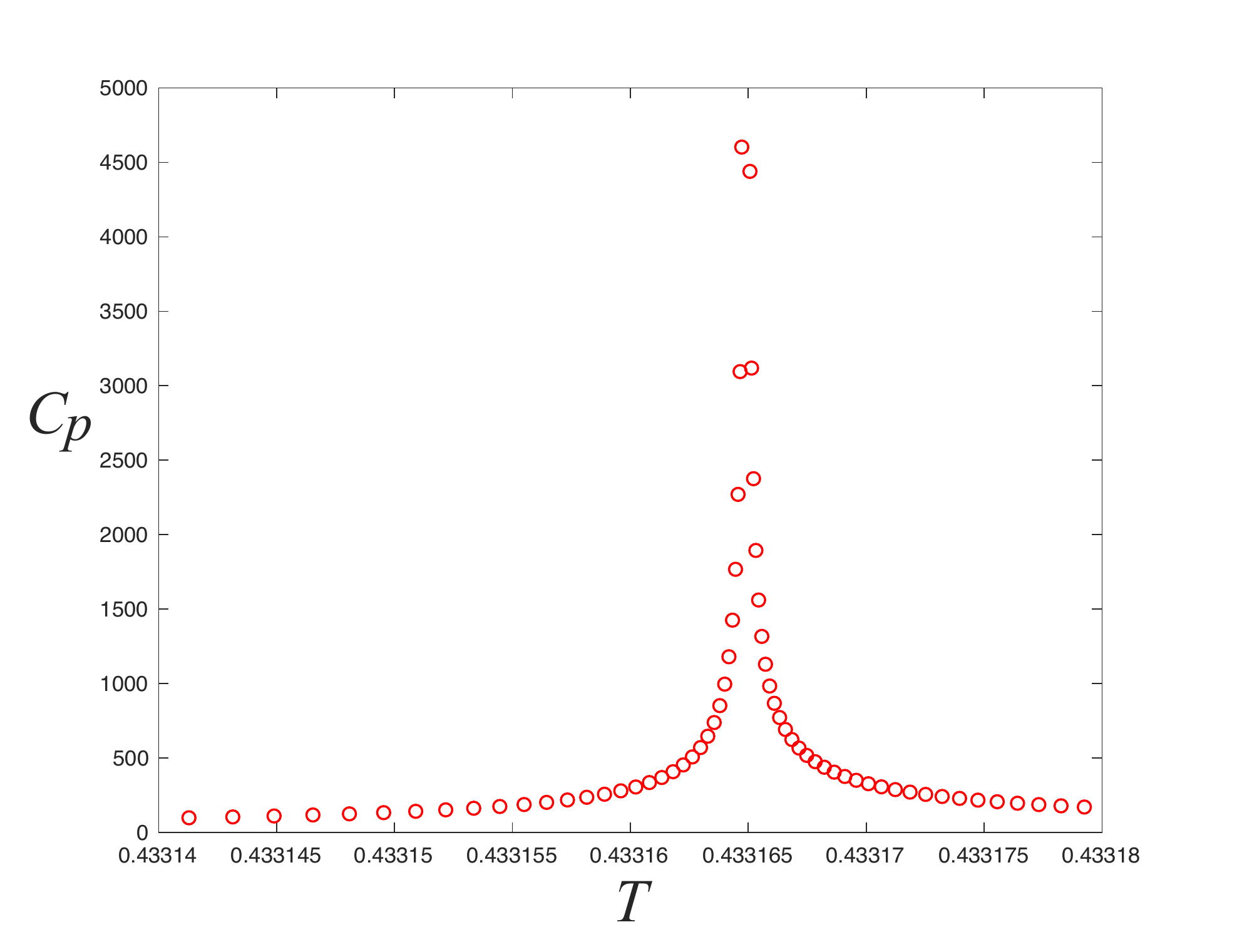}
\caption{\label{fig:critical-peak-in-Cp-RN} The 
divergence in 
$C_p(T)$ for the Reissner--Nordstr\"{o}m black hole, at  $p=p_c$ as~$T$ passes through~$T_c$.}
\end{wrapfigure}
In this paper it is the dependence on temperature that interests us, and so we should extract that. Using equation~(\ref{eq:Cp-4Q}) with the equation of state in~(\ref{eq:temperature-and-volume-RN}),  implies the function $C_p(T)$. Well away from the critical regime, the function  $C_p(T)$ grows from zero at  $T=0$ and behaves as $C_p=\pi T^2/2p^2$ as $T\to\infty$.  The large $T$ behaviour (with $C_V=0$) is in fact a characteristic ``ideal gas'' behaviour\cite{Johnson:2015ekr} that governs all the solutions we will consider in this paper, and also  the Schwarzschild solution. It follows from the fact that the equation of state (and hence all the thermodynamic functions) become dominated by the highest power of $R_+$, rendering all other contributions (from charge and rotation) subleading: $2pR_+=T$. Near the critical regime, the function $C_P(T)$ develops some structure, arising from the zeros in the denominator of equation~(\ref{eq:Cp-4Q}). These zeros are simply the turning points of the equation of state equation. The first order transitions occur before those zeros appear (if they are isolated) but  they merge at  the second order critical point and  $C_p(T)$ shows a classic divergence (known to have the van der Waals critical exponent\cite{Kubiznak:2012wp}). This is shown (for $Q=0.1$) in figure~\ref{fig:critical-peak-in-Cp-RN}.

\section{Kerr--AdS Black Holes}
\label{sec:kerr-black-holes}

For our first exhibit with $C_V\neq0$, let us turn to the Kerr--AdS spacetime, which has metric:
\begin{eqnarray}
ds^2&=&-\frac{\Delta}{\rho^2}\left(dt-\frac{a\sin^2\theta}{\Xi}d\phi\right)^2 +\frac{\rho^2}{\Delta}dR^2+\frac{\rho^2}{\Delta_\theta}d\theta^2+\frac{\Delta_\theta\sin^2\theta}{\rho^2}\left(adt-\frac{R^2+a^2}{\Xi}\right)^2\ ,\nonumber\\
&&{\rm with}\quad\Delta\equiv \frac{(R^2+a^2)(\ell^2+R^2)}{\ell^2}-2mR\ ,\quad \Delta_\theta\equiv1-\frac{a^2}{\ell^2}\cos^2\theta\ , \nonumber\\
&& {\rm and}\quad \rho^2\equiv R^2+a^2\cos^2\theta\ , \quad \Xi\equiv1-\frac{a^2}{\ell^2}\ .
\end{eqnarray}
(Here we are working in four dimensions again, for clarity.)
The horizon at $R_+$, is again the largest solution of $\Delta(R_+)=0$. The physical mass and angular momentum are $M=m/\Xi^2$ and $J=aM$, respectively. The entropy is again a quarter of the area of the horizon, $S=\pi(R_+^2+a^2)/\Xi$, but the thermodynamic volume turns out to be independent of $S$\cite{Cvetic:2010jb}. It is hard to write the equation of state $T(p,V)$ in closed form, but the mass, and a bit of algebra, yields enthalpy  in the form $H(S,p,J)$, from which $T(S,p,J)$, and $V(S,p,J)$,  are readily derived\cite{Caldarelli:1999xj,Dolan:2011xt}: 
\begin{eqnarray}
H(S,p,J)&=& \frac{1}{2}\sqrt{\frac{\left(S+\frac{8pS^2}{3}\right)^2+4\pi^2\left(1+\frac{8pS}{3}\right)J^2}{\pi S}} \ ,\label{eq:kerr-enthalpy}\\
T(S,p,J)&=&\frac{1}{8\pi H}\left[\left(1+\frac{8pS}{3}\right)\left(1+8pS\right)-4\pi^2\left(\frac{J}{S}\right)^2\right] \ ,\label{eq:kerr-temp} \\
V(S,p,J)&=&\frac{2}{3\pi H}\left[S\left(S+\frac{8pS^2}{3}\right)+2\pi^2 J^2\right]\ .\label{eq:kerr-volume}
\end{eqnarray}
This form is enough for our purposes. For a fixed value of $J$ one can readily extract from equation~(\ref{eq:kerr-temp}) the behaviour $T(S)$ at fixed pressure slices, and from there compute $C_p(T)$ along that isobar at will. 
With only a little more effort, using equation~(\ref{eq:kerr-volume}) one can fix a volume and deduce, within a specified domain in the $(p,S)$ plane, the coordinates of that isochoric curve. Those same coordinates can then be used in equation~(\ref{eq:kerr-temp}) to determine the temperatures at points along that curve. This determines $T(S)$ at fixed volume slices, and hence $C_V(T)$. Amusingly, Kerr--AdS has a family of isotherms that qualitatively resembles that discovered for Reissner--Nordstr\"{o}m--AdS, and apparently shares the van der Waals critical behaviour seen there\footnote{It was shown analytically (to quadratic order in $J$) in ref.\cite{Gunasekaran:2012dq}, that the system has the van der Waals critical exponent.}\cite{Dolan:2011xt}. 
 \begin{wrapfigure}{r}{0.5\textwidth}
\centering
\includegraphics[width=0.5\textwidth]{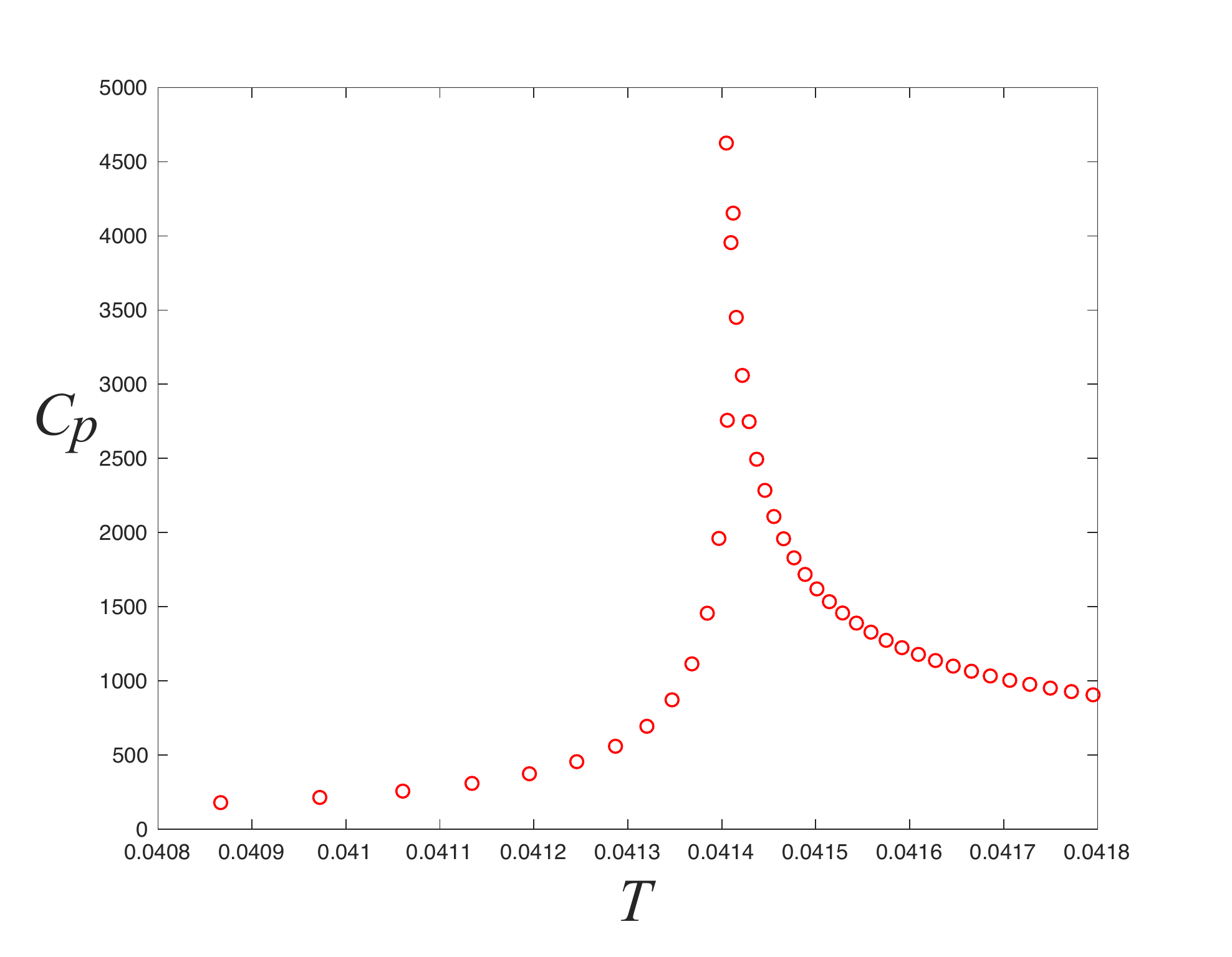}
\caption{\label{fig:critical-peak-in-Cp-kerr} The divergence in the specific heat $C_p(T)$ for the Kerr black hole, at  $p=p_c$ as $T$ passes through~$T_c$.}
\end{wrapfigure}
It is therefore natural to stay in the neighbourhood of the critical in examining our results for the specific heats. The critical point  is (for $J=1$)  at $T_c\simeq 0.0413$, $p_c\simeq0.00280$, $V_c\simeq122.39$.  Figure~\ref{fig:critical-peak-in-Cp-kerr} shows the case of $C_p(T)$ along the critical isobar, displaying the classic divergence.  There is already interesting new information in this function. Aside from the divergence, the marked rise in the specific heat  on the right side of it signals the presence of additional degrees of freedom in the system, as compared to the Reissner--Nordstr\"{o}m case ({\it c.f.} figure~\ref{fig:critical-peak-in-Cp-RN}). 

\begin{wrapfigure}{l}{0.48\textwidth}
\centering
\includegraphics[width=0.48\textwidth]{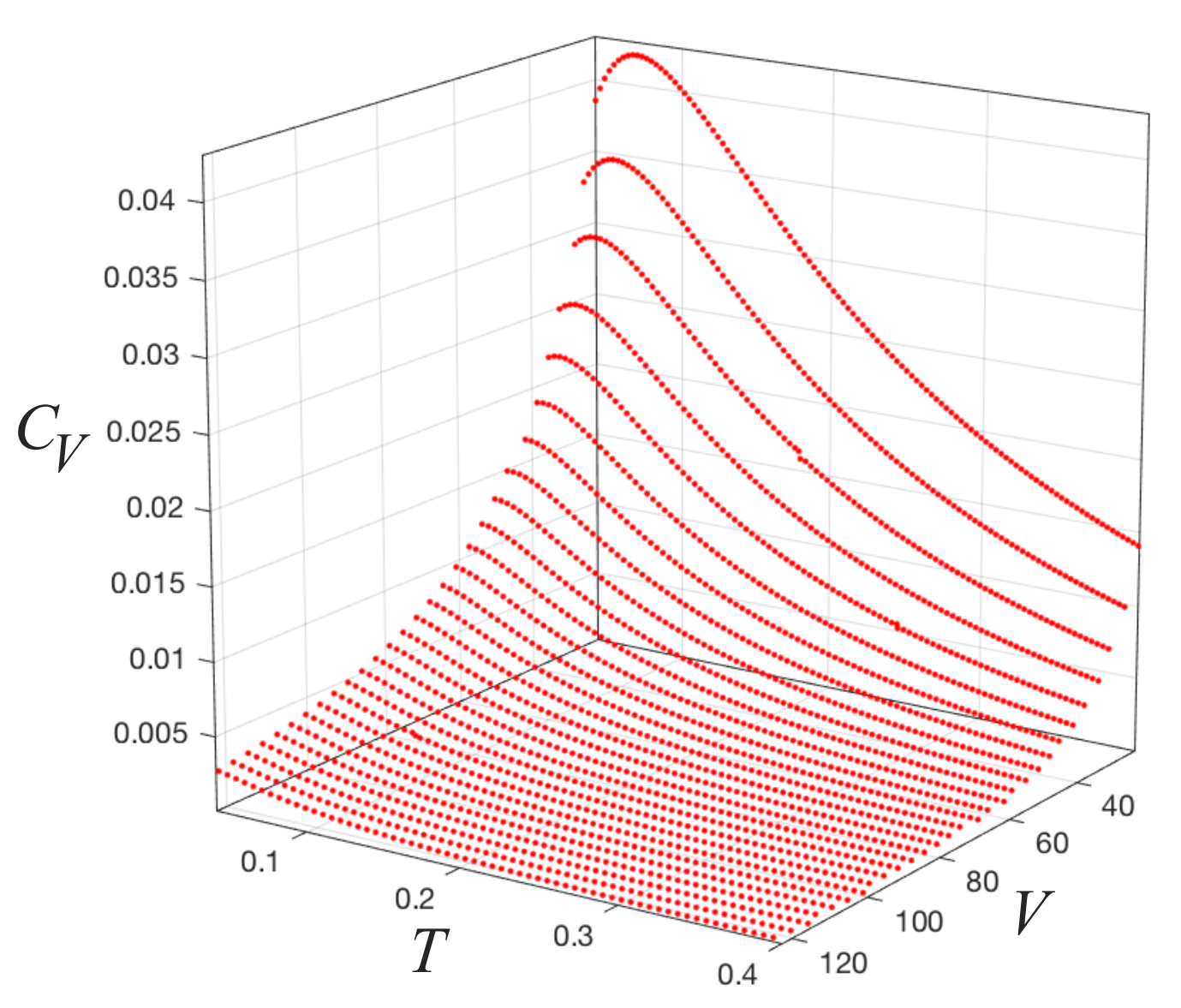}
\caption{\label{fig:crit-volume-Cv-kerr} The Schottky--like peaks in the specific heat $C_V(T)$ occurring at  $T>T_c$, for various  volumes.}
\end{wrapfigure}

 Things get even more interesting for $C_V(T)$,  slices of which are displayed in figure~\ref{fig:crit-volume-Cv-kerr}, ranging from $V\sim0.18 V_c$ to $V\sim V_c$.  There we see {\it e.g.} that for $V\sim21.95$ (the furthest curve) there is a distinctive peak in $C_V(T)$, at the value $C_V=0.0421$ at around $T\sim0.0659$, after which the function decays toward zero as the temperature increases. Two remarks are in order here. The first is that we are assured that $C_V(T)\to 0 $ as $T\to\infty$ since at large $T$ the $J$--dependent terms in all of the thermodynamic quantities are overwhelmed by the contributions from $S$ and $p$. In other words, the equation of state  reverts to the same ``ideal gas'' equation of state that controls Reissner--Nordstr\"{o}m and Schwarzschild at large $T$. So that there is some kind of a peak is assured. 
 
%
The second remark is that one might worry that there are subtleties coming from the  details of the first--order region below $T_c$ (the analogue of the region below the short--dashed (green) line in figure~\ref{fig:pv-diagram}), but notice that the round top of the peak is clearly visible at a temperature {\it above} $T_c$, and so that region cannot affect the conclusion that the round peak is a physically accessible feature here. Figure~\ref{fig:crit-volume-Cv-kerr} and further exploration shows that for other choices of fixed volume, as one reduces $T$, a portion of the peak  is {\it always} swept out. Whether one gets the round peak top and starts to decrease again simply depends upon the choice of volume. In some cases, before cresting the curve one reaches a temperature at which one begins to enter the first order region. Thereafter, the discussion becomes more complex since the unstable ($C_p<0$) region is entered. Formally one can  continue to maintain that fixed volume and continue to lower $T$ in an analogue of supercooling, and continue to trace out the curve. Indeed, general thermodynamic considerations assure us that $C_V(T)\to0$ as $T\to0$, completing the other side of the peak. 

It is interesting to ponder the location of the  peak, and the round top, when it appears, is as good a marker as any. Note that  in  figure~\ref{fig:crit-volume-Cv-kerr}, when $V=V_c$ (the closest curve), the top of  the peak does not appear for $T>T_c$ and not so for higher $V$. Put differently, the round top of the peak that seems to be accessible is just toward the upper left of the critical point's position in the  $p{-}V$ plane. It is hard to write an analytic expression for its position, but on general grounds one might expect that it is not independent of the critical point's position. The only free parameter here is $J$, and so once the critical point's position is fixed, there's nothing left over to freely move the position of the peak, and so they must be related. Dimensional analysis shows that the critical point (and  hence the accompanying Schottky--like peak) scales with $J$ as $p_c\sim 1/J, V_c\sim J^{3/2}$, with $T_c\sim 1/J^{1/2}$ and  $S_c\sim J$.

\section{STU Black Holes}
\label{sec:stu-black-holes}

 The next examples we will explore are  the so--called ``STU'' black holes (asymptotically AdS versions). Again, the thermodynamic volume for these was worked out in the literature\cite{Caceres:2015vsa}, and seen to be an independent function from  the entropy. Therefore $C_V\neq0$ again, and it seems prudent to study its temperature dependence. Cutting to the chase, the result will be that we again have the peaked behaviour, as might be anticipated by our work in the previous section, and especially the fact that these black holes have the same large $T$ asymptotics that gives vanishing $C_V(T)$ at $T\to\infty$. Moreover, the round top of the peak can again sometimes be seen at temperatures above where the first order critical region begins.
 
Actually,  the Reissner--Nordstr\"{o}m black hole is a special symmetric case of an STU black hole, as we will recall shortly. This leads to a nice picture: choices that break the symmetry to give the more general STU holes with the additional degrees of freedom are  akin to a sort of ``doping'' process. Degrees of freedom (resulting in a finite energy window) are  added to the system resulting in $C_V\neq0$.  This is exactly analogous to the sort of doping one might do in a material, resulting in Schottky peaks in the experimental data.

\noindent The STU--AdS metric is (again, working in four dimensions for presentational clarity)\cite{Sabra:1999ux,Duff:1999gh,Cvetic:1999xp}:
\begin{eqnarray}
	ds^2&=& -(H_1H_2H_3H_4)^{-1/2}f(r)dt^2+(H_1H_2H_3H_4)^{1/2}\left(f(r)^{-1}dr^2+r^2(d\theta^2+\sin^2\theta d\phi^2)\right)\ ,\nonumber\\
	 &&{\rm with}\quad f(r)\equiv1-\frac{2m}{r}+\frac{r^2}{\ell^2} H_1H_2H_3H_4\ ,
\end{eqnarray}
and  the functions $H_i=(1+q_i/r)$ are given in terms of four parameters $q_i$ ($i=1,\cdots,4$). The horizon is at $r=r_+$, where $r_+$ is given by $f(r_+)=0$. This equation determines the parameter $m$ in terms of $q_i$ and $r_+$:
\begin{equation}
\label{eq:m-equation}
m=\frac{r_+}{2}\left(1+\frac{r_+^2}{\ell^2}\prod_iH_i(r_+)\right)\ .
\end{equation}
Then the extended thermodynamics yields\cite{Caceres:2015vsa} the following expressions for the enthalpy $H$, as well as $T, S,V$, and $p$:
\begin{eqnarray}
        H &=& m+\frac14\sum_i q_i\ , \label{eq:enthalpy}\\
	T &=& \frac{f^\prime(r_+)}{4\pi}\prod_i H_i^{-1/2}(r_+)\ , \label{eq:temperature}\\
	S &=& \pi \prod_i \sqrt{r_++q_i}\ , \label{eq:entropy}\\
	V &=& \frac{\pi}{3}r_+^3\prod_i H_i(r_+)\sum_j H_j^{-1}(r_+)\ , \label{eq:volume}\\
	p &=& \frac{3}{8\pi G\ell^2}\ . \label{eq:pressure}
\end{eqnarray}
The parameters $q_i$ are related to a family of four physical charges that are given by
   \begin{equation}
   Q_i=\frac{1}{2}\sqrt{q_i(q_i+2m)}\ .
   \label{eq:fixedcharge}
   \end{equation}
   There is a natural (perhaps even helpful) eleven dimensional picture of all this\cite{Chamblin:1999tk,Cvetic:1999rb,Cvetic:1999xp}. AdS$_4$ naturally arises from M--theory as a decoupling limit of an M2--brane, which as a two--dimensional object in 11~dimensions, as 8 transverse spatial directions. The decoupling limit yields the AdS$_4\times S^7$ background of 11~dimensional supergravity.  One could set the M2--brane spinning, if so desired. There are four orthogonal planes in which those spins can be independently chosen. These spins correspond, after Kaluza--Klein reduction, to  charges (our $Q_i$) under  four $U(1)$ gauge symmetries. (See ref.~\cite{Duff:1999rk} for a review.) 
   The four dimensional theory also has three scalars with a potential term (not explicitly displayed here since we won't need them), and the $U(1)$ gauge fields' actions have a coupling to the scalars. In the special case of all four charges being equal, the scalars decouple, and we return to the symmetric case of an Einstein--Maxwell--AdS action, and the solution is  Reissner--Nordstr\"{o}m--AdS, with $Q=\sum_i Q_i/4$.  Shifting the radial coordinate  to $R=r+q$ we obtain the metric in equation~(\ref{eq:RN-AdS-solution}), although $Q$ there is twice $Q$ here.
   
Treating, as we do here, the charges $Q_i$ as fixed, there is a rich thermodynamic phase structure for these black holes, generalizing the remarkable van der Waals like behaviour noticed in ref.~\cite{Chamblin:1999tk}. As observed in ref.~\cite{Caceres:2015vsa}, the broad distinguishing  features of this large family of solutions can be captured by dividing into four classes: Four equal charges (Reissner--Nordstr\"{o}m), which we will denote as 4-$Q$,   three equal charges (3-$Q)$, two equal charges (2-$Q$), and 1 charge (1-$Q)$. The 3-$Q$ case qualitatively has the same phases as the 4-$Q$ (Reissner-Nordstr\"{o}m) case, with a qualitatively similar family of isotherms as seen in figure~\ref{fig:pv-diagram}.     The 2-$Q$ case loses the van der Waals--like critical point: There is instead a critical temperature marking the change from isotherms with a single $C_p>0$ branch, to ones where $C_p$ changes sign once. Finally, the 1-$Q$ case is rather similar in some respects to the case of no charge at all: All the isotherms have $C_p$ going from negative to positive as volume increases.

We have seen that much can be extracted from the 4-$Q$ case quite readily since the thermodynamic quantities can be written in  closed form in terms of physical quantities. Unfortunately, the non--diagonal cases are extremely hard to write in closed form. This is because one must hold the physical charges $Q_i$ fixed, while the solution and various thermodynamic quantities are written in terms of the parameters $q_i$, which must be allowed to vary when computing the derivatives of the entropy to compute specific heats. 


As a result of this difficulty in writing closed form solutions, we instead employ numerical methods to study the equations of state of the STU black holes, as well as computing $C_p(T)$ and~$C_V(T)$.  The strategy we chose was to pick a region of interest in the $p{-}V$ plane and divide it into an $n\times n$ grid of points, doing the following computation for each point in the grid\cite{Caceres:2015vsa}: For a given~$V$,  equation~(\ref{eq:volume})  can be solved to give  $r_+(V,q)$. 

This solution $r_+(V,q)$ may be input into equation~(\ref{eq:m-equation}), to yield $m(p,V,q)$. Hence, we can determine $q(p,V, Q)$ in terms of the fixed~$Q$ by using this $m$ in equation~(\ref{eq:fixedcharge}). From this we can determine $T(p,V)$  and $S(p,V)$ using equations~(\ref{eq:temperature}) and~(\ref{eq:entropy}). The functions thus determined over the $p{-}V$ grid are stored as ``data'' sets for later mining.  Grids with $n=100$ were typically used. Both Maple and MatLab were used for generating and mining information.

\begin{wrapfigure}{r}{0.5\textwidth}
\centering
\includegraphics[width=0.5\textwidth]{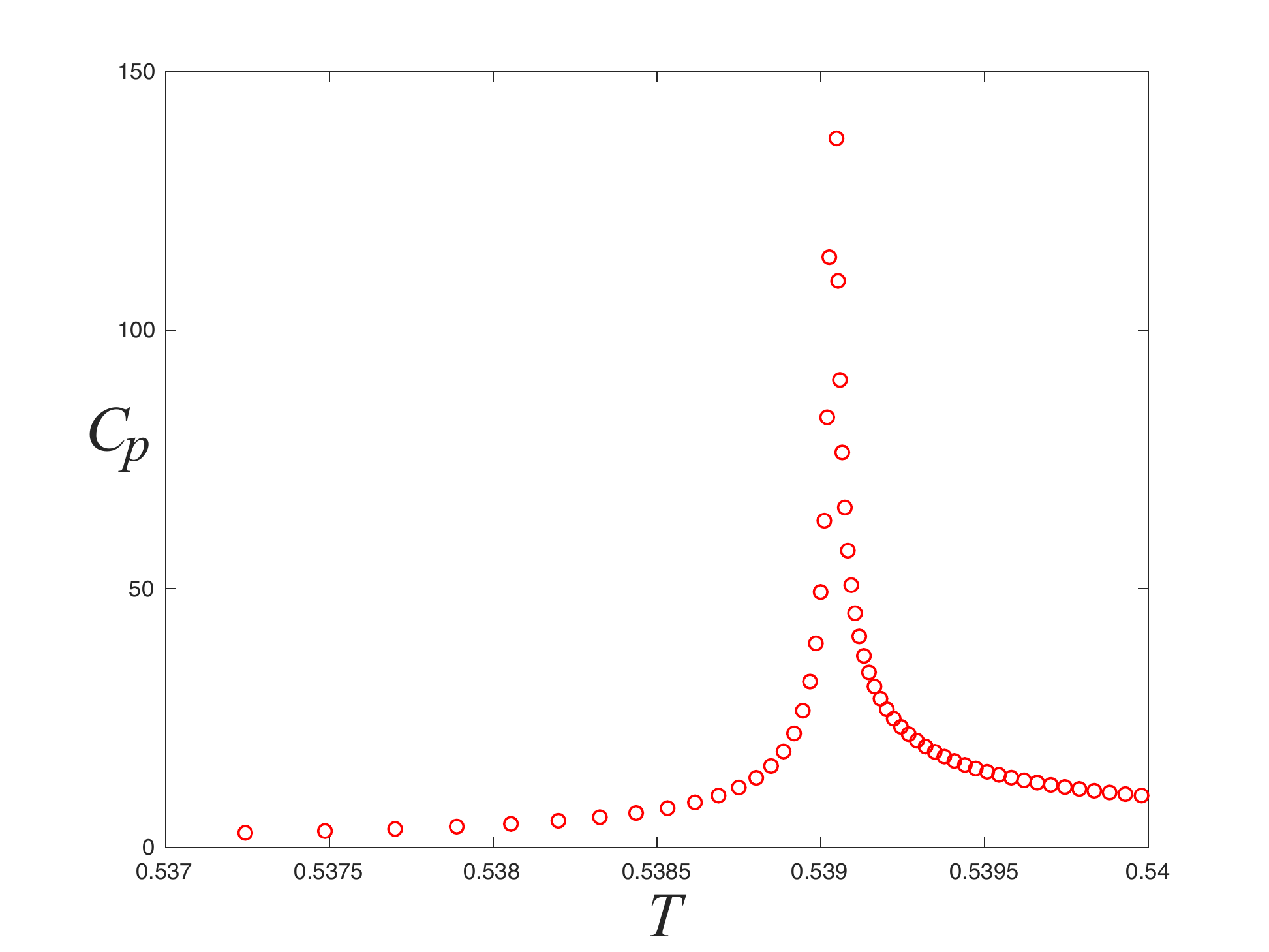}
\caption{\label{fig:critical-peak-in-Cp-STU} The divergence in the specific heat $C_p(T)$ at  $p=p_c$ as $T$ passes through $T_c$, for 3-$Q$ STU black holes.}
\end{wrapfigure}

With the points generated to make the functions $T(p,V)$ and $S(p,V)$, 
a numerical differentiation can be performed to generate the specific heats. Indeed for the 3-$Q$ case the function $C_p(p,V)$ is seen to have 
two ridges of divergences (the analogue of vanishing denominators giving the turning points of the isotherms that we saw before)  that meet, defining a second order critical point.  (For the 2-$Q$ and 1-$Q$ cases, $C_p(p,V)$ has a single ridge.) It is possible to use these data to extract the behaviour of the function $C_p(T)$ for some ranges of parameters and given choice of~$p$. We chose  $Q=0.05$ (see below). For 3-$Q$, picking the neighbourhood of the critical pressure $p=p_c\simeq 0.524$ yields the expected sharp divergence as $T$ passes through~$T_c\simeq 0.539$, shown in figure~\ref{fig:critical-peak-in-Cp-STU} ({\it c.f.} figures~\ref{fig:critical-peak-in-Cp-RN} and~\ref{fig:critical-peak-in-Cp-kerr})\footnote{Ref.\cite{Caceres:2015vsa} showed numerically that this divergence is again in the van der Waals universality class.}.

We now turn to $C_V$. We found good numerical control of our slices of parameter space if we kept $Q$ small. The particular value we chose was $Q=0.05$,  as a check that our methods were reproducing the results of ref.\cite{Caceres:2015vsa}. Significantly large $Q$ led to numerical problems rooted in the fact that for every grid point the system must numerically solve high order polynomial equations for $q$, and sometimes large regions of the $p{-}V$ plane proved too difficult.

So we worked with $Q=0.05$. Since $Q$ also controls the size of $C_V$ compared to $C_p$, this meant that $C_V$ was very small compared to $C_p$ and therefore the process of numerical differentiation was prone to significant numerical error. Since the features of~$C_V$ are our interest, it was worthwhile  proceeding with care. To this end, we used the fact that (from using $dS=\left({\partial S}/{\partial T}\right)_V dT+ \left({\partial S}/{\partial V} \right)_T dV$ and a Maxwell relation):
\begin{eqnarray}
\label{eq:CpCV}
	C_p-C_V=TV\alpha_p^2\kappa_T\ ,
\end{eqnarray}
where  $\alpha_p\equiv V^{-1} (\partial V/\partial T)_p$ is the isobaric thermal compressibility, $\kappa_T\equiv - V(\partial p/\partial V)_T$ is the isothermal bulk modulus. From this,  $C_V$ is more accurately computed numerically by working out its difference from $C_p$ using those more accurately computable derivatives.  This was tested in our numerics by first checking that, to the accuracy needed, the relation~(\ref{eq:CpCV}) gives zero for $C_V$ in the~4-$Q$ case (done numerically as a testbed). %

\begin{figure}[h]
\centering
\subfigure[]{\centering\includegraphics[width=0.48\textwidth]{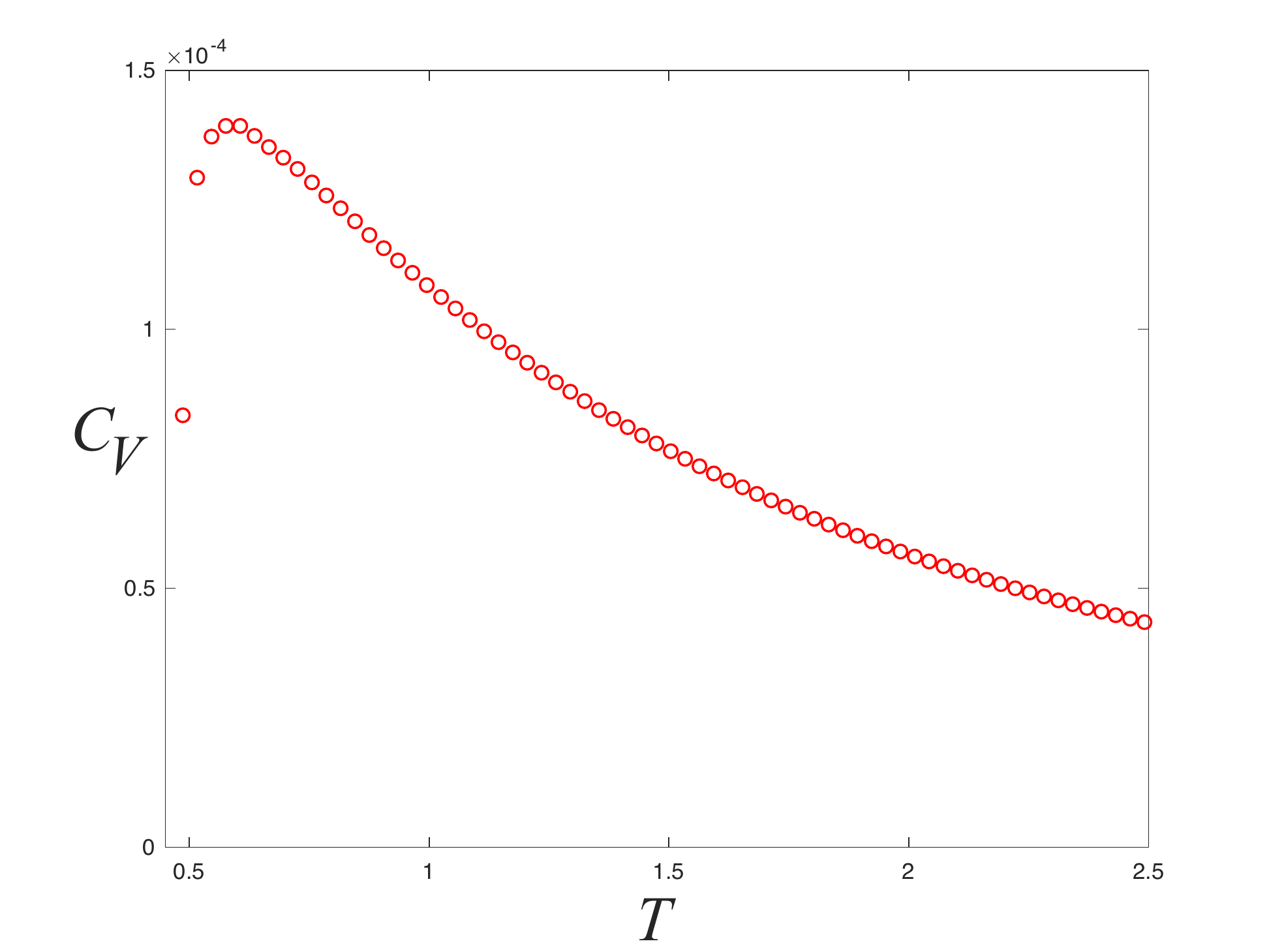}
\label{fig:crit-volume-Cv}}
\subfigure[]{\centering\includegraphics[width=0.46\textwidth]{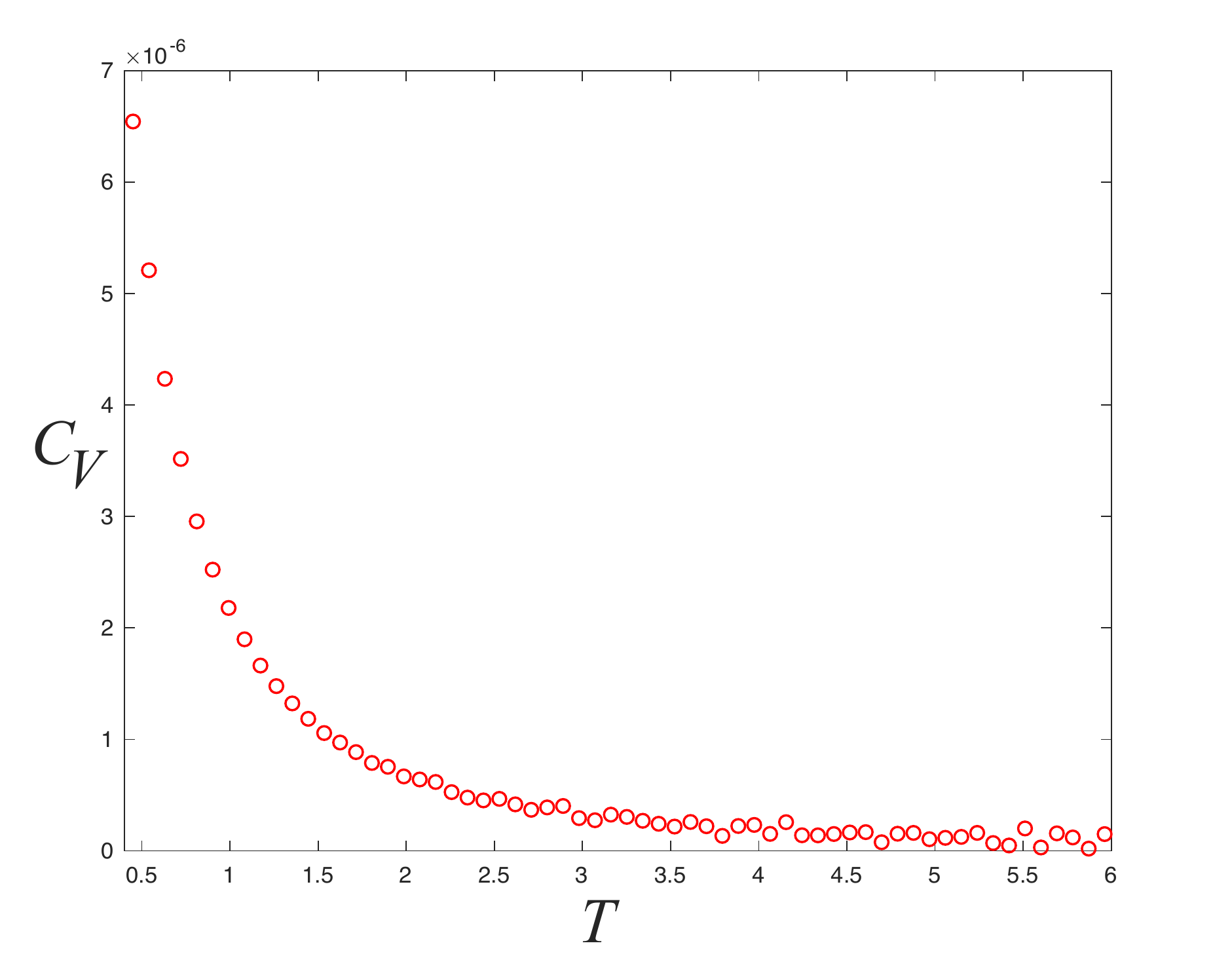}
\label{fig:higher-volume-Cv}}
\caption{(a) The specific heat $C_V(T)$ at   $V\simeq4.166V_c$, showing a peak somewhat above $T_c\simeq0.539$. (b) The specific heat $C_V(T)$ at   $V\simeq3.0$ over a wider range of temperatures.}
\end{figure}

Applying it next to the 3-$Q$ case yields a clear non--zero signal for $C_V$. As an example, a plot of $C_V(T)$ at a $V>V_c\simeq0.0264$ is shown in figure~\ref{fig:crit-volume-Cv}, showing a round peak for  $T>T_c\simeq0.539$, nothing remarkable happens to $C_V$, as might be expected.  It is of order $10^{-4}-10^{-5}$ smaller than $C_p$ in this region. 
Again there is a steady {\it decrease} with increasing temperature fitting with the fact that at large $T$ the whole system asymptotes to the  ``ideal gas'' limit mentioned earlier, where $C_p=\pi T^2/2p^2$ and $C_V=0$.   
Figure~\ref{fig:higher-volume-Cv} shows  similar behaviour at lower temperature and higher volume, where the overall function is another two orders of magnitude smaller (the limits of the numerical accuracy are beginning to show in the figure at higher temperatures, as the function drops to values consistent with zero. We observed similar behaviour  2-$Q$ and 1-$Q$. We shall refrain from presenting the plots here since they have no new features beyond what we saw for 3-$Q$. Again, since we have  for $T\to\infty$ that  $C_V\to0$, we deduce as before that there  is at least one maximum in the function $C_V(T)$ over the whole temperature range. How deeply one can go into the low~$T$ region (mapping out the full peak)  depends again on the choice of fixed volume, and at what temperature one enters the critical region. We did not explore the detailed crossover region since this would require the precise location of the first order transition, which needs a computation of the free energy of the solutions. That is an interesting project beyond the scope of this paper.

So overall, a similar pattern has emerged for the 3-$Q$ case as the one we saw for Kerr in the previous section. Again, the round peak is in the neighbourhood of the critical point, and  the lack of another free parameter beyond $Q$ suggests that its location is tied to that of the critical point, for which the locations scale according to $p_c\sim 1/Q^2, V_c\sim Q^3, T_c\sim 1/Q, S_c\sim Q^2$, as for the 4-$Q$ (Reissner--Nordstr\"{o}m) case.

\section{Discussion}
\label{sec:discussion}
Our central proposition is worth repeating here: Like any other thermodynamic system, an examination of the specific heat as a function of temperature can provide valuable clues as to the nature of the available underlying degrees of freedom. It is one of the oldest and most basic links between statistical mechanics and thermodynamics, and of course helped formed the foundations of our quantum theory of matter through key work of  Einstein, Debye, and others\cite{Einstein1907,Debye1912,rosser1982introduction,mandl1995statistical,kittel1996introduction}.  $C_V(T)$ is in a sense a cleaner diagnostic tool than $C_p(T)$, since the latter also accounts for energy given up to changing the volume of the system as well as exciting the degrees of freedom.

Now we turn to black holes, embedded in a theory of quantum gravity, a theory for which we are still struggling to understand the basic degrees of freedom. In this context they are thermodynamic objects, and the extended thermodynamics affords them with the opportunity to have both a $C_p$ and a $C_V$. In light of the above, it is striking that $C_V=0$ for Schwarzschild, Reissner--Nordstr\"{o}m, and other static pure gravity solutions. The natural conclusion is that  (despite the fact that their phase structure so often resembles that of models of familiar interacting matter) {\it they do not have the same kinds of degrees of freedom at all}, since $C_V(T)$ should track degrees of freedom that  can be excited without volume changes. A corollary of this is that the degrees of freedom in play (determining the overall phase structure, etc) can only be excited if also producing a volume change. This is of course very reminiscent of phenomena in  basic string theory, and it is natural to conjecture that this is not an accident, especially given that (at fixed $p$) there is a dual description in terms of thermalized open strings connecting branes (giving the gauge theory at low energy). This suggests that a general lesson is simply that extended thermodynamics  distinguishes between degrees of freedom that are stringy in origin and those that are not, such as momentum or Kaluza--Klein states, the latter only contributing to $C_V$. This is worth further exploration.

Moving forward and carefully studying two types of  black hole for which $C_V\neq0$ yielded a striking result. The function $C_V(T)$  for various choices of $V$ possess analogues of Schottky peaks. Put differently, these black holes (in extended thermodynamics) have the more familiar ``matter--like" degrees of freedom, but they come with a restricted set of energy levels. These are reminiscent of impurities, defects, or hyperfine splitting effects familiar in the laboratory, resulting in a band of available energies, or a finite set of discrete energy levels. 

It would be interesting to better understand the origin of these special degrees of freedom. Perhaps they have a natural description in some underlying string or M--theory description. The spinning brane origins of the STU black holes might provide a clue. In this regard if is worth noticing that for STU black holes in asymptotically flat space, there is a surprising connection between the $T=0$ entropy formula and a measure of entanglement  between three qubits\cite{Duff:2006uz}. There is a natural way of building these black holes in type~IIB string theory out of four sets of D3--branes wrapped on $T^6$ that makes the basis of discrete level choices quite explicit\cite{Borsten:2008ur}. (See ref.~\cite{Borsten:2012fx} for a review of this subject.) The qubits don't seem to play a dynamical role in the physics (so far), and of course the  details are all quite different from how AdS--STU black holes are realized, but it would be interesting to see if there could be lessons to be learned. Perhaps in the underlying string/M description of AdS--STU black holes there could be something that yields our finite energy window. 

There is certainly more to be done to improve the exploration of the systems that we have considered here. The numerics can probably be approached more carefully in order to get better control of more of parameter space and hence fully map out the peaks. Moreover, there are possibly more details to be understood about how the peak is embedded in the 2-$Q$ and 1-$Q$ cases, and how much of the round top can be seen. Our studies there were not exhaustive, not the least because the precise location of the first order lines have not been worked out in the literature (this is also true for the 3-$Q$ case). It would also be interesting to determine what sets the location of the Schottky--like peak in the 1-$Q$ case, where no critical point is present. 

Of course, there are other black hole solutions that are worth exploring from the perspective of this paper, in order to learn the nature of the degrees of freedom implicated by their specific heats. Most interestingly, it would be of interest to find a black hole solution for which $C_V(T)$ might be extracted analytically, or at least in as closed a form as for the $C_p(T)$ of Reissner--Nordstr\"{o}m of section~\ref{sec:rn-black-holes}. That might allow for clearer statements to be made about the degrees of freedom that contribute to the peak.

Finally, regardless of the origin of these states that contribute so interestingly to $C_V(T)$,  their presence and nature demand that  they be put to work\footnote{The mild pun is entirely accidental.}. The context is likely to be firmly in the realm of extended thermodynamics, since it is at constant volume that they are isolated from the other degrees of freedom. There could be applications to wherever black holes are of use, and this includes (but is not limited to) holographic studies of field theory and quantum information. They could also be useful for further illuminating the nature of quantum gravity.

Aside from those subjects, extended thermodynamics can be regarded\cite{Johnson:2017hxu,Johnson:2017ood} as a useful toolbox supplying new kinds of equations of state (organized through  solutions of Einstein's equations) that can be helpful in the study of a broad class of thermodynamic systems, some of which even have experimental realizations. Our more refined understanding of the degrees of freedom that contribute to the physics could help in creating useful models. We hope to report on   application examples shortly\cite{metoappear}.

 
\section*{Acknowledgements}
The work of CVJ  was funded by the US Department of Energy  under grant DE-SC 0011687.  CVJ would  like to thank  the Aspen Center for Physics for hospitality, and  Amelia for her support and patience.

\bibliographystyle{utphys}
\bibliography{black_hole_specific_heats}

\end{document}